%% file: articlePST.tex
\documentclass[11pt,a4paper,twosides]{article}

\usepackage{url}

\usepackage{tabularx}
\usepackage{rotating}

\usepackage{graphicx}
\usepackage{amsmath}
\usepackage{amssymb}
\usepackage{bbding}           
\usepackage{hhline,colortbl}  

\usepackage{tikz}
\usetikzlibrary{arrows,decorations.pathreplacing,shapes,fit,backgrounds,shadows}

\usepackage{booktabs}          

\usepackage{xcolor}            

\input{texdefinitions.inc}

\begin{document}

\def\Title{\MTR{}: A Meta-Model for Information Flow in Trust and Reputation Architectures}
\title{\Title}

\author{
Eugen \textsc{Staab} \\
netRapid GmbH \& Co. KG (\url{http://www.netrapid.de/}) \\
\texttt{eugen.staab@netrapid.de}
\and
Guillaume \textsc{Muller}\\ 
Work partly done at: \\
Escola Polit\'{e}cnica de S\~{a}o Paulo (\url{http://www.pcs.usp.br/~lti/}) and \\
Commissariat \`{a} l'\'{E}nergie Atomique (\url{http://www-list.cea.fr/}) \\
\texttt{guillaume.muller.pro@gmail.com}
}

\date{August 12, 2010}

\maketitle

\begin{abstract}
  We propose \MTR{}, a meta-model for the information flow in
  (computational) trust and reputation architectures.
  On an abstract level, \MTR{} describes the information flow as it is
  inherent in prominent trust and reputation models from the
  literature.
  We use \MTR{} to provide a structured comparison of these models.
  This makes it possible to get a clear overview of the complex
  research area.
  Furthermore, by doing so, we identify interesting new approaches for
  trust and reputation modeling that so far have not been
  investigated. \\
\end{abstract}

\input{introduction.inc}

\input{definitions.inc}

\input{big_picture.inc}

\input{organising_existing_models.inc}

\input{related_works.inc}

\input{conclusion.inc}

\bibliographystyle{amsalpha}
\bibliography{references}

\end{document}

%% file: texdefinitions.inc
\def\check{$\checkmark$}
\def\cross{}

\definecolor{mydarkgrey}{gray}{0.5} 
\definecolor{mylightgrey}{gray}{1} 

\def\obs#1#2{\text{obs}_{#1}^{\left<#2\right>}}   
\def\eva#1#2{\text{eval}_{#1}^{\left<#2\right>}}  
\def\tb#1#2{\text{tb}_{#1}^{#2}}                  
\def\ti#1#2{\text{ti}_{#1}^{#2}}                  
\def\rc#1#2#3{\left(#1\right)_{[#2]}^{#3}}        

\def\observationp{observation}

\def\MTR{$\mathbb{M}\text{ITRA}$} 

\newcommand{\otoprule}{\midrule[\heavyrulewidth]}

\def\keywords#1{\textbf{Keywords:} #1.}

%% file: introduction.inc
\section{Introduction}
\label{Section:Introduction}

Open and decentralized systems are vulnerable to buggy or malicious
agents. To make these systems robust against malfunction, agents need
the ability to assess the reliability and attitudes of other agents in
order to choose trustworthy interaction partners.  To this end, a
multitude of trust and reputation models have been and still are being
proposed in the literature.
In short, each such model generally defines all or some of the three
following processes: how evidence about the trustworthiness of an
agent is gathered, how this evidence is combined into a final
assessment, and how this final assessment is used in decision-making.

Currently, it is difficult to get an overview of what has been done in
the area, and what needs to be done.  The main reasons for this are:
that the proposed trust and reputation models use no common
terminology;
that they are not compatible in their basic structure; and that their
respective contributions are evaluated against different metrics.
While not considering the last point in this article, the first two
questions motivate the introduction of an abstract model that allows
researchers to organize their models in a unified way.


To this end, we propose \MTR{}, a meta-model for the information flow
in computational trust and reputation architectures.  \MTR{}
formalizes and organizes the flow of information inside and between
agents.  More precisely, it describes the top-level processes to
gather evidence, and to combine it with information exchanged with
other agents.  \MTR{} makes use of four simple concepts of information
processing, namely the \emph{observation}, the \emph{evaluation}, the
\emph{fusion} and the \emph{filtering} of information, and abstracts
away from numerical computations.  Although being an abstract model,
\MTR{} captures important concepts used by existing trust and
reputation models.  This way, \MTR{} provides a big-picture of the
trust and reputation domain and paves the way for a structured survey
of the domain.

The model is useful for the community in at least four respects.
First, it serves as a terminological and structural framework to
describe new models.  Secondly, it provides a means for researchers to
classify and compare existing approaches in this domain.  In addition
to this, \MTR{} helps to identify new approaches to model trust and
reputation, as we will show in this article.  Finally, it helps
newcomers to get a concise overview on the structure of computational
models of trust and reputation.

The remainder of the article is organized as follows.  In
Sect.~\ref{sec:basicConcepts}, we introduce basic concepts of trust
and reputation modeling.  We describe the \MTR{} model in
Sect.~\ref{Section:The big picture}.  Following this, in
Sect.~\ref{Section:Organizing}, we use \MTR{} to classify existing
models and identify what has not yet been done in the research field
of trust and reputation.  Finally, we review related work in
Sect.~\ref{Section:RelatedWorks} and we draw conclusions in
Sect.~\ref{Section:Conclusion}.

%% file: definitions.inc
\section{Basic Concepts}
\label{sec:basicConcepts}

In this section, we describe basic concepts and notations that are used in this article.

\subsection{Trust Beliefs/Intentions/Acts and Reputation}
\label{concept:trust_beliefs}

\input{schema_reasoning.tex}

Following~\cite{McKnight2001ClassesOfTrust}, we distinguish three
concepts that are often confused in the literature: the \emph{trust
  belief}, the \emph{trust intention} and the \emph{trust act}.
In the same spirit as for the BDI architecture~\cite{Rao1996BDI}, raw
observations are at first evaluated and then used to form trust
beliefs, which in turn, are used to build trust intentions. Finally,
these intentions can be used as one criterion in the decision-making
process, eventually leading to a trust or distrust act.
Figure~\ref{fig:short_sequence} provides a schematic view of the trust
information chain.

A basic trust belief, which we denote with $\tb{\theta}{\Gamma}$,
reflects the view of an individual agent $\theta$ on the
trustworthiness of the agents $\Gamma$.  In the common case, the set
of agents $\Gamma$ contains only one agent.  However, trust beliefs
can also reflect how an agent $\theta$ thinks about a group of agents
(see also~\cite{hales02groupReputation,falcone08categories}); the
agents belonging to the group $\Gamma$ have to be similar in some
regard and so, experiences with any of them may to some extent be
evidence for the trustworthiness of the group as a whole. For example,
consider several agents being employed by a certain company; in such a
situation, the agents can be judged in their roles as employees of
this company, and their characteristics can be ``generalized'' to
other agents in the same company~\cite{falcone08categories}.

There is a second type of trust beliefs, which is also called
``reputation'': a collective assessment of a group of agents about
other agents ~\cite{WordNet08Reputation}.  Here, the corresponding
trust beliefs are the estimate \emph{by an individual} of what could
be such a shared opinion of a group of agents $\Theta$ about other
agents $\Gamma$.  We denote this collective trust believe by
$\tb{\Theta}{\Gamma}$.

Each trust belief is relating to a certain context, for which it is
valid.  The context includes a particular task, and the environmental
conditions, under which the trustee is believed (or not) to
successfully carry out this task on behalf of the trustor.  We will
discuss this concept of context in more detail later.  Now, for given
trustor(s) $\Theta$, trustee(s) $\Gamma$ and a certain context, there
can only be one unique trust belief.  This makes sure that the trustor
cannot believe that trustee(s) $\Gamma$ are at the same time
trustworthy \emph{and} untrustworthy concerning a context.
Nevertheless, the various kinds of trust beliefs can still be
contradictory.  For example, an agent can believe that the agents that
belong to a certain group are usually untrustworthy, but that a
specific agent in this group is well known and believed to be
trustworthy.  The reasoning about how to deal with such situations is
typically what occurs in another process, that does not fall into the
scope of this paper.

While trust beliefs are solely estimates about the trustworthiness of
other agents, the process of forming \emph{trust intentions}
incorporates also strategic considerations or characteristics of the
trustee.  For example, although a trustee believes another agent to be
trustworthy, he might still be very pessimistic about relying on this
agent.  When forming trust intentions, an agent actually transforms
trust beliefs, which are based on the past behavior of other agents,
into its own intended future behavior towards them. This typically
corresponds to computing the ``shadow of the
future''~\cite{Axelrod84ShadowOfFuture}. Trust intentions of an agent
$\alpha$ towards an agent $\gamma$, derived from sets of trust
beliefs, are denoted by $\ti{\alpha}{\gamma}$.

\subsection{Acquisition of Evidence}

Two types of information exist that can be used as input for a trust
reasoning process.  An agent can make direct observations about and
evaluations of the behavior of other agents, and it can also receive
messages from other agents that contain observations, evaluations or
trust beliefs.

\subsubsection{Direct Observations and Evaluations}

A trust belief about an agent $\gamma$ is derived from information
that provides evidence for $\gamma$'s trustworthiness.  We call such
evidence an \emph{evaluation} (see
Fig.~\ref{fig:short_sequence}). 
An evaluation is a subjective interpretation of a set of \emph{direct
  observations} about $\gamma$'s behavior towards some other agent
$\delta$.  In other words, the process of evaluation decides whether
the observations are evidence for trustworthiness of $\gamma$.
We write $\eva{\alpha}{\delta,\gamma}$ to denote an evaluation that is
done by agent $\alpha$ concerning the behavior of agent $\gamma$
towards agent $\delta$.  Analogously, the direct observations made by
$\alpha$ on an interaction between $\gamma$ and $\delta$ are denoted
by $\obs{\alpha}{\delta,\gamma}$.

\subsubsection{Communicated Evidence}

Besides directly acquired information, an agent can use information
received by messages from other agents.
We introduce the notation $\rc{x}{\theta,\dots,\beta}{\alpha}$ for a
message $x$ that is sent by agent $\beta$ to agent $\alpha$ through a
path of transmitters $\beta$ (direct sender to $\alpha$), \dots,
$\theta$ (initial sender). We call such a message \emph{communicated
  evidence}.
To ensure the correctness of the indicated path, either the final
receiver can assess the correctness of this chain of transmitters
(e.g., by spot-checking agents on the path and asking them whether
they have sent the message as is), or mechanisms are put in place to
prove that the message has indeed taken the indicated path and which
intermediary has made which modifications (by means of cryptography,
e.g., public-key infrastructure and signing).

Direct observations that an agent makes on its own can only be
incorrect if the sensors with which the observations were made are
faulty. Communicated evidence additionally can be wrong when the
sender is dishonest or incompetent. 

\subsection{Context and Uncertainty}
\label{subsec:context_uncertainty}

Each observation, evaluation, trust belief or trust intention is in
general only valuable if two pieces of information are attached to it:
\begin{enumerate}
  \item the context to which the information
  applies; 
  \item a measure of uncertainty of the information
  itself. 
\end{enumerate}

The concept of \emph{context} is a vital part for evidence-based trust
reasoning.  \cite{mcknight} give the example that ``one would trust
one's doctor to diagnose and treat one's illness, but would generally
not trust the doctor to fly one on a commercial airplane''.
If some direct observations are made in a certain context, then this
context should be annotated also to the evaluations (and the trust
beliefs and trust intentions) that are based on these observations.
As a result, context annotations should be propagated through the
information chain shown in Fig. \ref{fig:short_sequence}.

A measure of \emph{uncertainty} is needed to express how uncertain a
piece of information is believed to be.  This is especially important
when a trust model incorporates communicated evidence that may be
biased or wrong.
Similarly to the context information, uncertainty should be propagated
through the whole information chain.

%% file: schema_reasoning.tex
\begin{figure}[h]
  \centering
  \begin{tikzpicture}[
    line/.style={thick,>=stealth'},
    oval/.style={shape=rectangle,rounded corners=2pt,very thick,draw=black!50,fill=white,drop shadow},
    ]
    \node[oval] (obs) {observations (\emph{obs})};
    \node[oval,below of=obs] (eval) {evaluations (\emph{eval})};    
    \node[oval,below of=eval] (tb) {trust beliefs (\emph{tb})};        
    \node[oval,below of=tb] (ti) {trust intentions (\emph{ti})};    
    \node[oval,below of=ti] (ta) {trust acts (\emph{ta})};    
    \draw[line,->] (obs) -- (eval);
    \draw[line,->] (eval) -- (tb);
    \draw[line,->] (tb) -- (ti);
    \draw[line,->] (ti) -- (ta);
  \end{tikzpicture}
  \caption{Information Chain in Trust Reasoning.}
  \label{fig:short_sequence}
\end{figure}
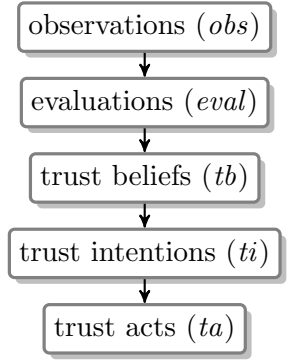

%% file: big_picture.inc
\section{The Meta-Model}
\label{Section:The big picture}

\begin{figure*}[t!]
 \begin{center}
   \includegraphics[width=\textwidth]{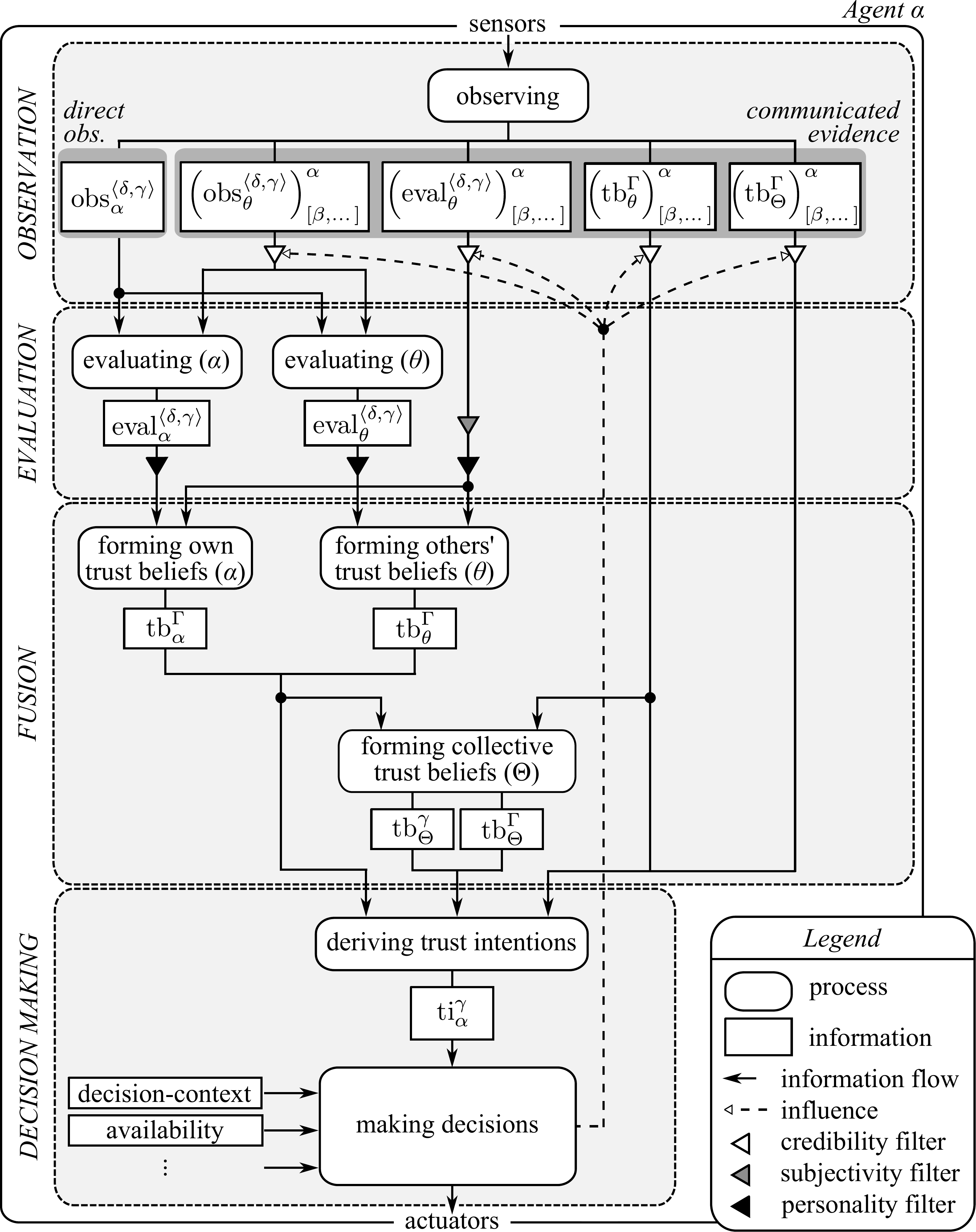} 
   \caption{Structure of \MTR{} for an agent $\alpha$. For the sake of clarity what is not shown in this illustration: agent $\alpha$ can decide to send any piece of information occurring in the illustration to other agents. \label{Figure:Overview}}
 \end{center}
\end{figure*}

In this section we present \MTR{}, illustrated in
Fig.~\ref{Figure:Overview}.  This meta-model organizes the different
ways of how the information chain of Fig.~\ref{fig:short_sequence} can
be realized in an agent $\alpha$'s trust model.  The agent can send
all pieces of information that occur in this information chain (the
square boxes in the Figure) to other agents; for the sake of clarity,
we did not illustrate such actions in Fig.~\ref{Figure:Overview}.

On the top level, \MTR{} divides the trust modeling process into four
consecutive sub-processes:
\begin{enumerate}
  \item observation;
  \item evaluation;
  \item fusion;
  \item decision-making.
\end{enumerate}

Let us first exemplify these sub-processes from the perspective of an
agent $\alpha$, which reasons about its trust in an agent $\gamma$.
In the \emph{observation} process, agent $\alpha$ tries to collect any
form of evidence it can get to assess $\gamma$'s trustworthiness.
Agent $\alpha$ judges the direct observations (either its own or the
communicated ones) about $\gamma$'s behavior in the \emph{evaluation}
process.
In the \emph{fusion} process, $\alpha$ can use its own evaluations,
and the filtered evaluations received from other agents, to form its
trust beliefs about $\gamma$ and an image of other agents' trust
beliefs about $\gamma$.
Finally, in the \emph{decision-making} process, $\alpha$ builds its
trust intentions and applies them in the respective situations.
Below, we describe each sub-process in greater detail and then have a
closer look at the issue of context-sensitivity.

\subsection{Observation}
\label{Section:Observation}

During \observationp{}, agent~$\alpha$ uses its sensors to capture
information about other agents and the environment.  In network
settings for instance, sensors could be network cards, that can
receive or overhear packets from the network.  In the general case,
the process of direct observation is \emph{subjective}, since
different agents may use sensors that differ in certain respects, for
instance in quality.

As mentioned earlier, information can result from direct observation
of the environment
or be received as 
communicated evidence.  For an observation
$\obs{\alpha}{\delta,\gamma}$, $\alpha$ may be the same agent as
$\delta$; in this case, $\alpha$ observes its own interaction with
some other agent $\gamma$, otherwise $\alpha$ is observing the
interactions of other agents.

Because communicated evidence is received from other agents, $\alpha$
needs to filter the information. Communicated evidence might be
incorrect for different reasons~\cite{Conte02Book,Avegliano08AAMAS}.
First, the communicator can intentionally provide wrong information,
i.e., lie.  For instance, the communicator might want to increase its
own reputation or the reputation of an acquaintance; this is usually
called ``misleading propaganda''.  It can also try to decrease the
reputation of another competing agent, which is usually called
``defamation''. Secondly, the communicator can provide information
that is not wrong but leads the receiver intentionally to wrong
conclusions.
For instance, it can hide information, give partial information or
give out-of-context information.  Finally, the communicator can
unwittingly communicate untrue facts.  Note that agent $\beta$, the
last agent who sent the information, needs not to be the originator of
the information~\cite{Conte02Book}, and therefore might alter the
information if it is not signed. The credibility filter should take
all these aspects into account and filter the information according to
how much the sources of this information are intended to be trusted,
more precisely, whether they are decided to be
trusted. 
In Figure \ref{Figure:Overview}, this is indicated by the
``influence''-arrows from ``making decisions'' to the
\emph{credibility filters}, which exactly try to filter out
information that does not seem to be credible.

\subsection{Evaluation}
\label{Section:Evaluation}

During the process of \emph{evaluation}, an agent evaluates sets of
direct observations about the behavior of other agents.
Such an evaluation estimates whether the set of observations provides
evidence for the agent in question being trustworthy or untrustworthy.
At this stage, it is not yet decided whether an agent is actually
believed to be trustworthy or not; sets of observations are only
examined for their significance in respect to an agent's
trustworthiness.
In many models, this is done by comparing the actual behavior of an
agent to what its behavior was expected to be like.
This expected behavior can, for instance, be determined by formal
contracts~\cite{Sabater02PhD} or social
norms~\cite{Castelfranchi00SocialOrder,Muller10LIAR}.  Various
representations are used to describe evaluations: binary
(e.g.,~\cite{Sen02BoolCase}), more fine-grained discrete (e.g.,
$\{-1,0,+1\}$ in eBay~\cite{eBaySite}), or continuous assessments
(e.g.,~\cite{Sabater02PhD,Huynh04FIRE,Muller10LIAR}).

As the sets of norms and established (implicit) contracts can be
subjective, an evaluation can be subjective too. Therefore, different
agents might contradictorily interpret the same observations as
evidence for trustworthy and untrustworthy behavior.
This is evident in the case of eBay~\cite{eBaySite} where each human
does the evaluation along his own criteria.  As a consequence, an
agent can try to emulate how other agents would evaluate observations,
to eventually emulate their trust beliefs.  Therefore, \MTR{} contains
two different ways of evaluation (see Fig.  \ref{Figure:Overview}):
using $\alpha$'s criteria, which results in evaluations
$\eva{\alpha}{\delta,\gamma}$; and the way $\alpha$ thinks $\theta$
would do the evaluation, which results in
$\eva{\theta}{\delta,\gamma}$.

For evaluation, it can be vital to know about the causality behind
what happened.  Indeed, if the causal relationships are not clear to
the evaluating agent, it can wrongly evaluate a failure to be evidence
for untrustworthiness, although it is actually an \emph{excusable}
failure -- or vice versa~\cite{castelfranchi00,Staab07Excusableness}.

\subsection{Fusion}
\label{Section:Computation}

In a third step, the different evaluations are \emph{fused} into trust
beliefs.
Trust beliefs can be formed based on evaluations of individual agents
or groups of agents, and can be about individuals or groups.
Figure~\ref{Figure:Overview} shows that an agent $\alpha$ can fuse
evaluations to get its \emph{own trust beliefs} $\tb{\alpha}{\Gamma}$;
or to emulate the trust modeling of another agent $\theta$, in order
to estimate this agent's \emph{$\theta$'s trust beliefs}
$\tb{\theta}{\Gamma}$.
In any case, if $\alpha$ uses evaluations $\eva{\beta}{\cdots}$
received from another agent $\beta$, it first needs to filter out the
``incompatible'' subjectivity inherent in the evaluations: For
evaluation, $\beta$ may have applied criteria that $\alpha$ does not
agree with, or -- when $\alpha$ is emulating the trust beliefs of
$\theta$ -- where $\alpha$ thinks that $\theta$ would not agree with.
In the figure, the check for the match of the applied criteria, and the potential adjustment of the evaluations, is named \emph{subjectivity filtering}.

The emulation of another agent $\theta$'s trust beliefs is for
instance needed when forming reputation, i.e., trust beliefs that a
certain group of agents $\Theta$ (with $\theta\in\Theta$) would
associate to a given trustee.
For this, also trust beliefs received from other agents can be
incorporated.
A subjectivity filter should not be applied to these received trust
beliefs, because reputation reflects the subjectivity of different
agents. Still, a credibility filter is applied to avoid a biased
reputation estimate.

Furthermore, if an agent forms trust beliefs based on interactions
where itself was not involved, it has to account for the relation
between the interacting agents.
Assume, an agent $\alpha$ receives from another agent the evaluation
$\eva{\theta}{\delta,\gamma}$, where all four agents $\alpha$,
$\theta$, $\delta$, and $\gamma$ are distinct agents.
If $\alpha$ wants to reason about $\delta$'s trustworthiness towards
itself, then it first needs to apply a \emph{personality filter}.
Here, information is filtered out that contains no evidence for the
behavior of $\delta$ towards $\alpha$, because it is specific for
interactions between $\delta$ and $\gamma$.
%
For example, imagine that the evaluation is about the behavior of a
mother ($\delta$) towards her child ($\gamma$), and she behaved very
trustworthy.  Then this behavior does not say much about how the
mother will behave towards another unrelated person (e.g., $\alpha$).
This shows that trustworthiness is \emph{directed}, and that this
direction has to be taken into account in trust reasoning.

Whenever fusing sets of evaluations or trust beliefs into a single
trust belief, it is particularly important to account for the
``correlated evidence'' problem~\cite{Pearl88Correlated}.  This
problem arises when different evaluations or trust beliefs are based
on the same observed interactions of agents. In other words, the
evidence expressed by the evaluations/trust beliefs ``overlaps''.  If
the reasoning agent is not aware of this overlapping, certain parts of
the evidence will wrongly be amplified and the resulting trust belief
be biased.

Many different approaches for fusing evaluations into trust beliefs
(e.g., \cite{singh,travos,reece07sla}), and trust beliefs into
community models (e.g., \cite{eigentrust,Schillo99WhoDealing}) have
been proposed in the literature.
Most importantly, each trust belief should incorporate the two
properties mentioned in Sect.  \ref{subsec:context_uncertainty}: the
uncertainty, i.e., how strong the belief is, and in which context(s)
it applies.

\subsection{Decision-Making}
\label{Section:DerivingTrustIntentions}

The last component in \MTR{} is the \emph{decision-making} process, which
consists of two steps:
\begin{enumerate}
  \item fix trust intentions based on trust beliefs;
  \item apply the trust intentions to make the final decision to act
  (or not) in trust.
\end{enumerate}

In the first step, a set of individual and/or collective trust beliefs
are used to \emph{derive} one or several trust intentions for
different contexts.
Trust beliefs can, for instance, be aggregated into trust intentions
by simply averaging over them, or by taking the most ``pessimistic''
or ``optimistic'' trust belief, etc.
But also, it is possible to ignore available negative trust beliefs
about another agent $\gamma$, and to decide to act in trust with
$\gamma$ in order to give this agent the opportunity to rethink its
behavior~\cite{Castelfranchi04AAMAS}.  This ``advance'' in trust,
which can in some cases be \emph{forgiveness}, accounts for the
dynamics of trust such as ``trust begets trust''
\cite{bradach89trustBegetsTrust}.

A trust intention $\ti{\alpha}{\gamma}$ has to reflect two things: in
which situations $\alpha$ actually intends to act in trust with
$\gamma$ (context), and how strong the intention is (uncertainty).
Trust intentions with these two properties can be used in
\emph{decision-making} in the same way as other criteria.
Although many trust and reputation models from the literature do not
separate the derivation of trust intentions from the process of
decision-making, we strongly argue for a separation of the two
processes. The reason is that the derivation of trust intentions is
specific to trust research, while a trust intention plays in decision
making the role of a context-sensitive criterion in the same way as
many other criteria (like availability of a potential partner). This
problem is however studied in depth in a research area called Multiple
Criteria Decision Making (MCDM) \cite{kaliszewski2006} and is not
specific to trust.

\subsection{Context in \MTR{}}
\label{Section:Context}

In general, \emph{everything} that can impact the behavior of a
trustee and is not part of the trustee itself, is said to belong to
the context~\cite{Dey00PhD}.
The more of the available context information is considered by a trust
and reputation model, the better the final decision can be.
We propose to arrange the different facets of context into five
classes:
\begin{enumerate}
  \item time (points in time or time intervals);
  \item external conditions:
  \begin{enumerate}
    \item physical conditions, 
    \item laws/norms,
    \item other agents nearby;
  \end{enumerate}
  \item type of the delegated task;
  \item contract;  
  \item information source.
\end{enumerate}

The \emph{information source} is the one that provides the
information, on which the reasoning is based, e.g., a sensor or
another agent.  This context facet is a special case, because it is not
linked to a trustee's behavior.  It is important though, since it
allows an agent to have several trust beliefs about the same trustee,
based on different information sources.

Two different contexts can be distinguished: the one in which the
information was taken, and the one in which a decision is made
(\emph{decision context}).
To fuse information that comes from different contexts, or use it in
decision-making for different contexts, an agent needs to know how
similar the different contexts are.  In the literature, many ways for
representing context information, and similarities between contexts,
have been proposed.  \cite{kinateder03architecture} represent context
relations in form of a weighted directed graph.
\cite{sensoy07ontology} use ontologies and a set of rules to represent
context information.
In a more general way, \cite{rehak07trust} represent context
information as points in a multi-dimensional space, where each
dimension represents one characteristic of the context (e.g., the
point in time, the dollar exchange rate, etc.). The distance between
two points in the context space, which is determined by some distance
metric, states the similarity of the two contexts.
These approaches have in common that an agent was given some
similarity metric about different contexts in advance.
\MTR{} does not assume that, and so, if the similarity metric cannot
be known in advance, an agent needs to learn the metric.

At which stage information that belongs to different contexts is
eventually fused, is up to the concrete model.  However, since during
this fusion process some information is lost, it should be done as
late as possible in the information chain. As a consequence, that
principle should generally be used when processing information.

%% file: organising_existing_models.inc
\section{Organizing Existing Models}
\label{Section:Organizing}

\input{table_observation.tex}

\input{table_evaluation_fusion_decision.tex}

In this section, we exemplify how existing trust and reputation (T\&R)
models can be classified by means of \MTR{}.  The approach we take
here is to investigate which types of data are used in the considered
trust models.  For a small selection of trust models, this is shown in
tables \ref{tab:ExistingModelsClassifiedWithMTR1} and
\ref{tab:ExistingModelsClassifiedWithMTR2}.  Table
\ref{tab:ExistingModelsClassifiedWithMTR1} lists types of data that
belong to the observation process.  Table
\ref{tab:ExistingModelsClassifiedWithMTR2} lists the remaining data
types, i.e. those in the evaluation, fusion and decision-making
processes.  A check-mark indicates whether a trust model accounts for
the respective kind of data.  Still, not every model accounts for the
context and uncertainty of the processed information.

Additionally to this classification approach, T\&R models can also be
described by specifying:
\begin{enumerate}	
  \item which filters are applied,
  \item where uncertainty is considered, and 
  \item which facets of context are accounted for at which stage. 
\end{enumerate}

Since the focus of this article lies on the meta model \MTR{} itself,
we leave it to future work to provide an extensive and comprehensive
classification using \MTR{}.  However, as it can already be seen in
the resulting tables, a clear picture emerges of what has been done
(columns with check-marks) and what needs to be done in the T\&R
modeling domain -- for example, there is no model with no empty
columns, i.e., that uses all available types of information.
Furthermore, the column
$\text{eval}_{\theta}^{{}\left<\delta,\gamma\right>}$ is always empty.
None of the here considered trust models\footnote{To our knowledge
  there is no such trust model in the literature.} uses the
intermediate step of simulating the evaluation of another agent to
eventually form a collective trust belief (reputation).  However, we
believe that this is something what humans do regularly; for example
in form of questions like ``What would my mother think about his
behavior?'' or ``How would my best friend judge this agent's
behavior?''. This issue deserves more attention as it makes it
possible to evaluate the behavior of other agents in cases where an
own opinion on their behavior is lacking.

%% file: table_observation.tex
\begin{sidewaystable}
\centering	
\renewcommand{\thempfootnote}{\alph{mpfootnote}}
\renewcommand{\thefootnote}{\alph{footnote}}
	\caption{Classification with \MTR{} (Observation).}
		\label{tab:ExistingModelsClassifiedWithMTR1}
		\begin{tabular}{cccccc} 
			
\toprule
	  					& \multicolumn{5}{c}{\cellcolor{mylightgrey}Observation}\\ 
			        & \cellcolor{mylightgrey} $obs_\alpha^{<\delta,\gamma>}$
			        & \cellcolor{mylightgrey} {$\left(\text{obs}_{\theta}^{{}\left<\delta,\gamma\right>}\right)_{\left[ \beta,\dots \right]}^{\alpha}$}
			        & \cellcolor{mylightgrey} {$\left(\text{eval}_{\theta}^{{}\left<\delta,\gamma\right>}\right)_{\left[ \beta,\dots \right]}^{\alpha}$}
			        & \cellcolor{mylightgrey} $\left(\text{tb}_{\theta}^{\Gamma}\right)_{\left[ \beta,\dots \right]}^{\alpha}$
			        & \cellcolor{mylightgrey} $\left(\text{tb}_{\Theta}^{\Gamma}\right)_{\left[ \beta,\dots \right]}^{\alpha}$ \\
		\otoprule
		
		   Marsh~\cite{Marsh94PhD}
			       & \cross{}     
			       & \cross{}     
			       & \check{}     
			       & \cross{} 
			       & \cross{}     
			       \\	\hline	
		
						Schillo et al.~\cite{Schillo99WhoDealing}  						
			       & \cross{}     
			       & \cross{}     
			       & \cross{}     
			       & \check{}\footnotemark[2]      
			       & \cross{}     
			       \\	\hline	

		   {\scshape Histos}~\cite{Zacharia1999SporasHistos}  
			       & \cross{}     
			       & \cross{}     
			       & \cross{}     
			       & \check{}\footnotemark[2]      
			       & \cross{}     
			       \\	\hline

		   {\scshape Sporas}~\cite{Zacharia1999SporasHistos}  
			       & \cross{}     
			       & \cross{}     
			       & \check{}     
			       & \cross{}     
			       & \cross{}     
			       \\	\hline
			{\scshape Beta-Rep. System}~\cite{josang02beta}  
			       & \check{}     
			       & \cross{}     
			       & \check{}\footnotemark[3]     
			       & \cross{}     
			       & \check{}\footnotemark[2]\footnotemark[4]     
			       \\	\hline

		   Sen and Sajja~\cite{Sen02BoolCase}  
			       & \cross{}     
			       & \cross{}     
			       & \check{}     
			       & \cross{}     
			       & \cross{}     
			       \\	\hline	

			{\scshape EigenTrust}~\cite{kamvar03eigentrust}
			       & \check{}     
			       & \cross{}     
			       & \cross{}     
			       & \cross{}     
			       & \check{}\footnotemark[2]      
			       \\ \hline
%

		   {\scshape Regret}~\cite{sabater03PhD}  
			       & \cross{}     
			       & \cross{}     
			       & \cross{}     
			       & \check{}\footnotemark[2]     
			       & \check{}\footnotemark[2]      
			       \\	\hline	
			            	       
			  {\scshape Secure}~\cite{cahill03trust}
			       & \check{}     
			       & \cross{}     
			       & \check{}     
			       & \check{}\footnotemark[2]     
			       & \check{}\footnotemark[1]\footnotemark[2]    
			       \\ \hline      
			 
			 Wang and Vassileva~\cite{wang03bayesian}
						 & \check{}     
			       & \cross{}     
			       & \cross{} 
			       & \check{}\footnotemark[2]    
			       & \cross{}     
			       \\	\hline	
			       
			 {\scshape PeerTrust}~\cite{xiong04peer}
						 & \check{}     
			       & \cross{}     
			       & \check{} 
			       & \cross{}    
			       & \cross{}     
			       \\	\hline
			       				       
			Capra and Musolesi~\cite{capra06pervasive}
			       & \check{}     
			       & \cross{}     
			       & \cross{}     
			       & \cross{}     
			       & \cross{}     
			       \\ \hline       
			       
			{\scshape Travos}~\cite{teacy06travos}
			       & \check{}     
			       & \cross{}     
			       & \check{}     
			       & \cross{}     
			       & \cross{}     
			       \\ \hline
	       
%
%
%
			Reece et al. \cite{reece07sla} 
			       & \check{}     
			       & \check{}     
			       & \cross{}     
			       & \cross{}     
			       & \cross{}     
			       \\ \hline
%

						{\c{S}}ensoy and Yolum \cite{sensoy07ontology}  
						 & \check{}     
			       & \check{}     
			       & \cross{} 
			       & \cross{}     
			       & \cross{}     
			       \\	\hline

			Wang and Singh~\cite{singh}
			       & \check{}\footnotemark[1]
			       & \cross{}     
			       & \cross{}     
			       & \cross{}     
			       & \cross{}     
			       \\ \hline		
%
			Falcone and Castelfranchi~\cite{falcone08categories}  
			       & \check{}     
			       & \cross{}     
			       & \cross{}     
			       & \cross{}     
			       & \cross{}     
			       \\ \hline

		   {\scshape Liar}~\cite{Muller10LIAR}  
			       & \check{}     
			       & \check{}     
			       & \check{}     
			       & \check{}\footnotemark[2]      
			       & \check{}\footnotemark[2]      
			       \\ \hline	
			 
			 Vogiatzis et al.~\cite{vogiatzis10probTrust}
			       & \check{}     
			       & \cross{}     
			       & \check{}      
			       & \cross{}       
			       & \cross{}       
			       \\			       
			\bottomrule
		\end{tabular}
\footnotetext[1]{implicitly}
\footnotetext[2]{only considering the case of $|\Gamma|=1$}
\footnotetext[3]{submitted to a central rating center}
\footnotetext[4]{provided by a central rating center}
\end{sidewaystable}

%% file: table_evaluation_fusion_decision.tex

\begin{sidewaystable}
	\centering	
\renewcommand{\thempfootnote}{\alph{mpfootnote}}
\renewcommand{\thefootnote}{\alph{footnote}}
	\caption{Classification with \MTR{} (Evaluation, Fusion, Decision-Making).}
	\label{tab:ExistingModelsClassifiedWithMTR2}
		\begin{tabular}{cccccccc}
		
		\toprule

			      & \multicolumn{2}{c}{\cellcolor{mylightgrey}Evaluation}
					 	& \multicolumn{3}{c}{\cellcolor{mylightgrey}Fusion}
						& \cellcolor{mylightgrey} Decision-Making\\

              & \cellcolor{mylightgrey} $\text{eval}_{\alpha}^{{}\left<\delta,\gamma\right>}$ 
			        & \cellcolor{mylightgrey} $\text{eval}_{\theta}^{{}\left<\delta,\gamma\right>}$
			        
			        & \cellcolor{mylightgrey} $\text{tb}_{\alpha}^{\Gamma}$
			        & \cellcolor{mylightgrey} $\text{tb}_{\theta}^{\Gamma}$
				      & \cellcolor{mylightgrey} $\text{tb}_{\Theta}^{\Gamma}$
			        
			        & \cellcolor{mylightgrey} $\text{ti}_{\alpha}^{\gamma}$ \\

		\otoprule

		  Marsh~\cite{Marsh94PhD}
			       & \check{}     
			       & \cross{}     
			       & \check{}\footnotemark[1]     
			       & \cross{}     
			       & \cross{}     
			       & \check{}     
			       \\ \hline
	
		  Schillo et al.~\cite{Schillo99WhoDealing}
			       & \check{}     
			       & \cross{}     
			       & \cross{}     
			       & \cross{}     
			       & \check{}\footnotemark[1]     
			       & \cross{}     
			       \\ \hline
			       	
			       				       %
		  {\scshape Histos}~\cite{Zacharia1999SporasHistos}
			       & \cross{}     
			       & \cross{}     
			       & \check{}\footnotemark[1]     
			       & \check{}\footnotemark[1]     
			       & \cross{}     
			       & \cross{}     
			       \\ \hline
			       
		  {\scshape Sporas}~\cite{Zacharia1999SporasHistos}
			       & \check{}     
			       & \cross{}     
			       & \cross{}     
			       & \cross{}     
			       & \check{}\footnotemark[1]     
			       & \cross{}     
			       \\ \hline
			       	
			{\scshape Beta-Rep. System}~\cite{josang02beta} 
			       & \check{}     
			       & \cross{}     
			       & \cross{}     
			       & \cross{}     
			       & \check{}\footnotemark[1]\footnotemark[2]     
			       & \cross{}     
			       \\ \hline		       
			       	
		  Sen and Sajja~\cite{Sen02BoolCase}
			       & \check{}     
			       & \cross{}     
			       & \check{}\footnotemark[1]     
			       & \cross{}     
			       & \cross{}     
			       & \cross{}     
			       \\ \hline		       
			       	
			{\scshape EigenTrust}~\cite{kamvar03eigentrust}
			       & \check{}     
			       & \cross{}     
			       & \cross{}     
			       & \cross{}     
			       & \check{}\footnotemark[1]     
			       & \cross{}     
			       \\ \hline

		  {\scshape Regret}~\cite{sabater03PhD}
			       & \check{}     
			       & \cross{}     
			       & \check{}\footnotemark[1]     
			       & \cross{}     
			       & \check{}\footnotemark[1]     
			       & \check{}     
			       \\ \hline
			         
			{\scshape Secure}~\cite{cahill03trust}
			       & \check{}     
			       & \cross{}     
			       & \check{}\footnotemark[1]     
			       & \cross{}     
			       & \cross{}     
			       & \check{}     
			       \\ \hline
			Wang and Vassileva~\cite{wang03bayesian}
       			 & \check{}     
			       & \cross{}     
			       & \check{}\footnotemark[1]     
			       & \cross{}     
			       & \cross{}     
			       & \check{}     
			       \\\hline
			       
			{\scshape PeerTrust}~\cite{xiong04peer}
			       & \check{}     
			       & \cross{}     
			       & \check{}\footnotemark[1]     
			       & \cross{}     
			       & \cross{}     
			       & \check{}     
			       \\\hline
			       
			Capra and Musolesi~\cite{capra06pervasive}
       			 & \check{}     
			       & \cross{}     
			       & \check{}\footnotemark[1]     
			       & \cross{}     
			       & \cross{}     
			       & \cross{}     
			       \\\hline			       
			{\scshape Travos}~\cite{teacy06travos}
			       & \check{}     
			       & \cross{}     
			       & \check{}\footnotemark[1]     
			       & \cross{}     
			       & \cross{}     
			       & \cross{}     
			       \\ \hline
		  Reece et al. \cite{reece07sla} 
			       & \cross{}     
			       & \cross{}     
			       & \check{}\footnotemark[1]     
			       & \cross{}     
			       & \cross{}     
			       & \cross{}     
			       \\ \hline

		  {\c{S}}ensoy and Yolum~\cite{sensoy07ontology}  
			       & \check{}     
			       & \cross{}     
			       & \check{}\footnotemark[1]     
			       & \cross{}     
			       & \cross{}     
			       & \cross{}     
			       \\\hline
			       
			Wang and Singh~\cite{singh}
			       & \check{}     
			       & \cross{}     
			       & \check{}\footnotemark[1]     
			       & \cross{}     
			       & \cross{}     
			       & \cross{}     
			       \\ \hline
%
%
%
						Falcone and Castelfranchi~\cite{falcone08categories}  
			       & \check{}     
			       & \cross{}     
			       & \check{}     
			       & \cross{}     
			       & \cross{}     
			       & \cross{}     
			       \\	\hline
%
		  {\scshape Liar}~\cite{Muller10LIAR}
			       & \check{}     
			       & \cross{}     
			       & \check{}\footnotemark[1]     
			       & \check{}\footnotemark[1]     
			       & \check{}\footnotemark[1]     
			       & \check{}     
			       \\ \hline
				
				Vogiatzis et al.~\cite{vogiatzis10probTrust}
					   & \check{}     
			       & \cross{}     
			       & \check{}\footnotemark[1]     
			       & \cross{}     
			       & \cross{}     
			       & \cross{}     
			       \\ 
		
		\bottomrule
    \end{tabular}
\footnotetext[1]{only considering the case of $|\Gamma|=1$}
\footnotetext[2]{computation performed by central rating center}
\end{sidewaystable}

%% file: related_works.inc
\section{Related Work}
\label{Section:RelatedWorks}

Numerous survey or overview papers on trust and reputation models have
been published.  Many works start with a basic classification and then
enumerate and describe existing models
\cite{sabater05overview,artz07survey,josang07survey,ramchurn04trust,LikMui02Rp}. 
Opposed to that, we tried in our work to use a rigorous methodology
for the organization of the state of the art: extract the core
structure of prominent trust and reputation models, in order to get
clear and simple differentiation criteria for the models of the
literature.

\cite{kinateder05} proposed a generic trust model that integrates
several existing trust
models. 
They show how to map each of these models to their model.  However,
they focus on the fusion of evaluations, whereas we examined the
overall structure of trust and reputation modeling.

\cite{Casare05OntologyReputation} define a functional ontology of
reputation, that is used in different
systems~\cite{Vercouter07IJCAI,Nardin08WONTO} in order to help agents
using different trust and reputation models inter-operate. These
approaches try to cope with a situation where different models exist,
whereas we try to propose a unified meta-model of trust and
reputation.


%% file: conclusion.inc
\section{Conclusion}
\label{Section:Conclusion}

In this article, we presented \MTR{}, a meta-model for trust and
reputation.
The model structures many essential concepts found in the literature
on evidence-based T\&R models.
The simple and generic structure of \MTR{} makes it suitable both for
experts to organize their models in a common way, and for newcomers to
easily enter the domain.  Although there is the possibility that a
specific T\&R model does not comply with \MTR{}, the fact that the
latter was derived from the study of many existing models argues for
its comprehensiveness, and we believe that its modularity should make
it simple to modify in order to encompass new elements.

Finally, by using \MTR{}, we classified existing T\&R models from the
literature. This classification revealed which kinds of information
are commonly used by these models, and which kinds of information are
often neglected. In this way, we found that none of the considered
models actually tries to emulate another agent's evaluation of a
trustee; this emulation would make it possible to form the reputation
of an agent in a new way.